\documentclass[a4paper]{article}
\usepackage{graphicx}
\usepackage{tabularx}
\usepackage{amssymb,amsmath,bm}
\usepackage{booktabs}
\usepackage{color}
\usepackage{INTERSPEECH2021}

\title{MIMO Speech Compression and Enhancement Based on Convolutional Denoising Autoencoder}
\name{You-Jin Li$^{1,2}$, Syu-Siang Wang$^3$, Yu Tsao$^2$, and Borching Su$^1$}
\address{
  $^1$Graduate Institute of Communication Engineering, National Taiwan University, Taipei, Taiwan\\
  $^2$Research Center for Information Technology Innovation, Academia Sinica, Taipei, Taiwan\\
  $^3$Department of Electrical Engineering, Yuan Ze University, Taoyuan, Taiwan}
  
\email{d05942004@ntu.edu.tw, sypdbhee@citi.sinica.edu.tw, yu.tsao@citi.sinica.edu.tw, borching@ntu.edu.tw}

\begin{document}

\maketitle

\begin{abstract}
For speech-related applications in IoT environments, identifying effective methods to handle interference noises and compress the amount of data in transmissions is essential to achieve high-quality services. In this study, we propose a novel multi-input multi-output speech compression and enhancement (MIMO-SCE) system based on a convolutional denoising autoencoder (CDAE) model to simultaneously improve speech quality and reduce the dimensions of transmission data. Compared with conventional single-channel and multi-input single-output systems, MIMO systems can be employed in applications that handle multiple acoustic signals need to be handled. We investigated two CDAE models, a fully convolutional network (FCN) and a Sinc FCN, as the core models in MIMO systems. The experimental results confirm that the proposed MIMO-SCE framework effectively improves speech quality and intelligibility while reducing the amount of recording data by a factor of 7 for transmission.
\end{abstract}


\section{Introduction}

Multichannel speech enhancement (MCSE) and speech compression techniques aid several real-time speech communications  in an Internet of things system \cite{cheng2020mass}. Conventional MCSE systems with a multiple-input single-output (MISO) configuration suppress environmental noise from multiple noisy inputs to provide decent sound quality and intelligibility on the single-channel output side \cite{de2017adaptive,607754, van2013multi}. In general, most MCSE algorithms were derived using beamforming-based approaches \cite{hoshuyama1999robust,emura2018distortionless,7820704,gannot2001signal}, wherein either the spatial diversity of received signals or the maximum signal-to-noise ratio (SNR) criterion was exploited to apply a linear filter function and preserve the desired signal \cite{warsitz2007blind,kellermann2008beamforming}. Several attempts further combine deep learning (DL) with conventional beamforming-based MCSE to provide a robust transfer function and increase the capability of the system to handle non-stationary noise environments \cite{zhang2017deep,cohen2017combined,qian2018deep,chang2019mimo}. In addition to beamforming-based approaches, some studies have enhanced noisy recordings directly through DL models. For example, the work in \cite{tawara2019multi} used a denoising auto-encoder (DAE) model to suppress noise in the time domain to preserve the speech signal in a specified spatial direction. Our previous work \cite{liu2019multichannel} utilized a fully convolutional neural network (FCN) and Sinc FCN (SFCN) on MCSE to achieve decent speech quality and intelligibility in both subjective and objective tests. Notably, the Sinc layer \cite{ravanelli2018speaker} used in SFCN provides more meaningful filters to decompose the model inputs for the following FCN model.

In addition to the improved sound quality, multichannel inputs also increase bandwidth, power consumption, and hardware costs for signal transmission and storage. An effective acoustic signal compression method is required to reduce the amount of captured data \cite{spanias2006audio}. For acoustic signal compression, speech coding (SC) approaches are applied to remove the statistical redundancies or perceptual irrelevancies of input audio signals \cite{malvar1990lapped, hiwasaki2008g, he2015applying,ozerov2013coding, faller2001efficient, breebaart2005parametric}. Traditional SC approaches, such as sub-band coding \cite{sayood2017introduction} and code-excited linear prediction \cite{jage2016celp}, are derived by considering temporal properties to compress a single-channel speech signal. In contrast, multichannel SC approaches, such as spatial audio coding \cite{faller2003binaural, faller2005parametric}, and modified discrete cosine transform \cite{suresh2012mdct, suresh2012mdct, chen2010spatial}, are applied to encode input signals by considering both coherence and statistical differences across channels. Generally, some level of distortion can be observed in coded and restored speech signals and slightly degrade the speech quality and intelligibility accordingly. Recently, DL techniques have been introduced in signal compression algorithms to implement SC systems \cite{biswas2020audio}. In \cite{cite6} and \cite{cite7}, an utterance was first analyzed using deep neural networks to extract phonological and prosodic speech representations to build novel speech codecs. In \cite{cite8} and \cite{cite9}, speech spectra were encoded by a deep auto-encoder that was trained with identical input and output signals. The associated codecs were then derived from the output nodes of the middle hidden layer. Meanwhile, DL models have been used as post-filters to enhance coded speech \cite{cite10}, \cite{cite11}, and have exhibited decent speech quality.

In this study, we propose a multi-input multi-output speech compression and enhancement (MIMO-SCE) framework. The proposed framework is based on a convolutional DAE (CDAE) \cite{liu2019multichannel} model, comprising encoder and decoder parts. During training, the CDAE is trained to process noisy multichannel speech signals to generate enhanced signals. Thereafter, the encoder and decoder are separately placed at the edge and server, respectively. During testing, the encoder part transforms noisy multichannel speech inputs into bottleneck features with reduced dimensions. The encoded bottleneck features are then transmitted to the server and processed by the decoder to recover the enhanced multichannel speech signals. Two CDAE models were implemented for the MIMO-SCE framework: an FCN-based (termed MIMO-SCE(F)) and an SFCN-based (termed MIMO-SCE(S)). Experimental results show that MIMO-SCE(F) and MIMO-SCE(S) can effectively reduce multichannel acoustic data by a factor of seven while improving speech quality and intelligibility.

The remainder of this paper is organized as follows: A review of related works is presented in Section \ref{sec:rel}, and  the concepts and architectures of the proposed MIMO-SCE(F) and MIMO-SCE(S) models are discussed in Section \ref{sec:propos}. Section \ref{sec:exp} describes the experimental setup and results. Finally, the conclusions of the study are presented in Section \ref{sec:conc}. 

\section{Related Works}\label{sec:rel}
In this section, we first review MISO SE systems. Subsequently, we review two CDAE models: FCN and SFCN.

\subsection{MISO SE system}
For the multichannel noisy input $\mathbf{Y}=[\mathbf{y}_{1},\mathbf{y}_{2},\cdots,\mathbf{y}_{N}]$, where $N$ denotes the number of channels, the MISO SE system aims to generate an enhanced speech signal $\hat{\mathbf{x}}$, where $\hat{\mathbf{x}}=f_{\theta}(\mathbf{Y}$, and ${\theta}$ denotes the model parameters and is estimated by minimizing the difference between the generated speech $\hat{\mathbf{x}}$ and the clean reference. During the test, for a given noisy multichannel input, the MISO SE generates an enhanced single-channel output.

\begin{figure}[t]
\centering \centerline{
\includegraphics[width=0.55\textwidth,clip]{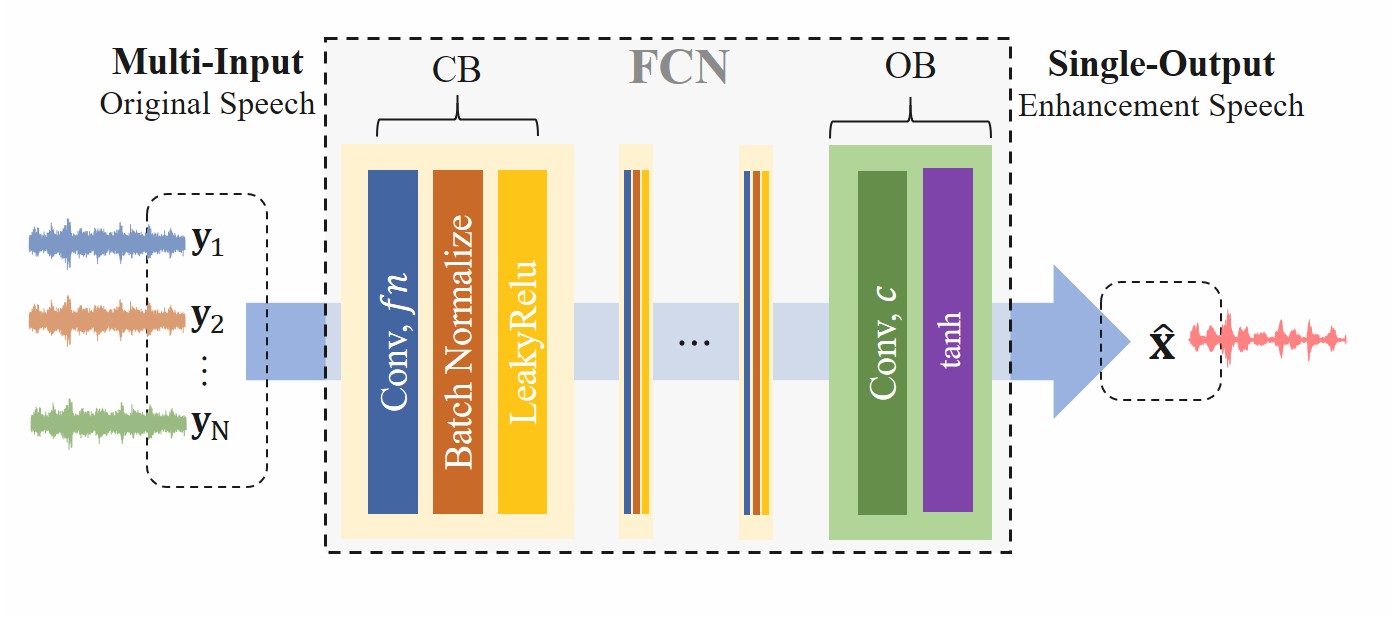}}
\caption{Architecture of MISO-FCN. CB and OB in the figure represent convolutional blocks and the output blocks, respectively.\vspace{-0.1cm}}
\label{fig:MI-FCN}
\end{figure}

\subsection{Two CDAE Models: FCN and SFCN}
In this study, two CDAE models were implemented. The first one is FCN, which comprises convolutional blocks (CBs), as shown in Fig. \ref{fig:MI-FCN}. Each CB consists of three components: convolution layer (Conv), batch normalization, and LeakyReLU. The filter number and filter length used in the convolution layer are $fn$ and  $fl$, respectively. A stack of CBs is concatenated for feature extraction and transformation. Finally, an output block (OB) consisting of a convolution layer and a tanh activation function is placed in the last part of the FCN. In an OB, the filter length of the convolution layer is defined as the output dimension $c$; for the MISO SE system, $c$ is equal to $1$. 

\begin{figure}[t]
\centering \centerline{
\includegraphics[width=0.55\textwidth,clip]{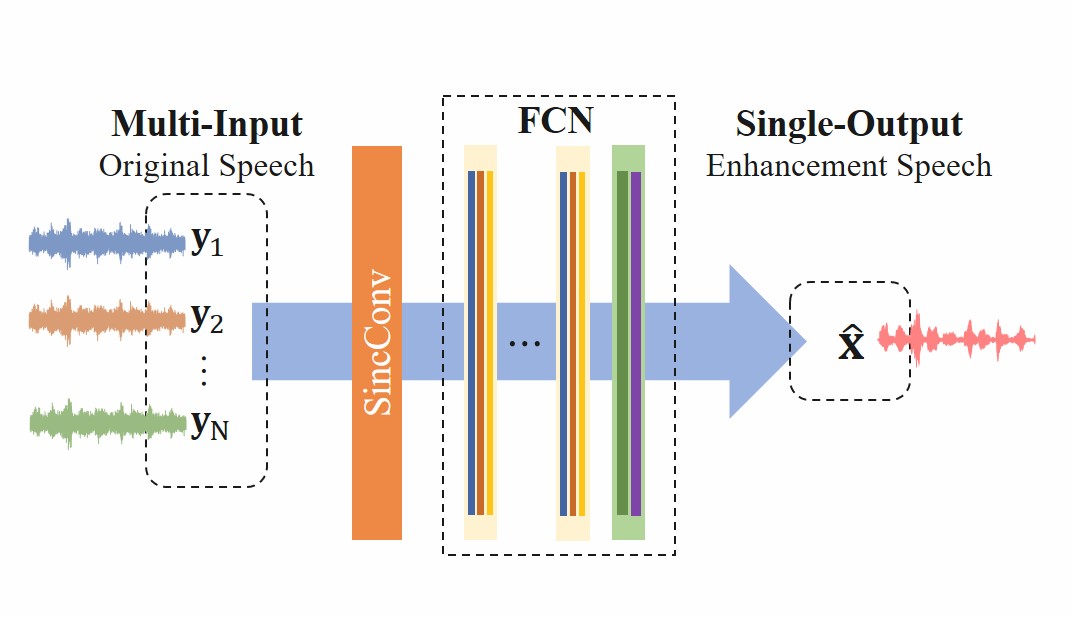}}
\caption{Architecture of MISO-SFCN\vspace{-0.1cm}}
\label{fig:MI-SFCN}
\end{figure} 

In our previous work \cite{liu2019multichannel}, we confirmed that SFCN can yield better MISO SE performance. The architecture of the SFCN is shown in Fig. \ref{fig:MI-SFCN}. The primary difference between the FCN and SFCN models is that SFCN adopts the Sinc convolution (SincConv) layer as the first CB. SincConv was designed and trained to provide various filter banks; thus, it can obtain band-pass information even for a limited amount and restricted diversity of training data. In addition, as SincConv contains fewer parameters, SFCN can be trained more efficiently.

\section{Proposed MIMO speech compression and enhancement framework}\label{sec:propos}
In this section, we introduce the architecture of the proposed MIMO-SCE framework. Two CDAE models are used as the core units to build the MIMO-SCE(F) and MIMO-SCE(S) systems. The goal of MIMO-SCE is to determine a function that transforms $\mathbf{Y}$ into multichannel clean speech signals, $\mathbf{X}$.

\begin{figure}[t]
{\centering
\includegraphics[width=0.55\textwidth,clip]{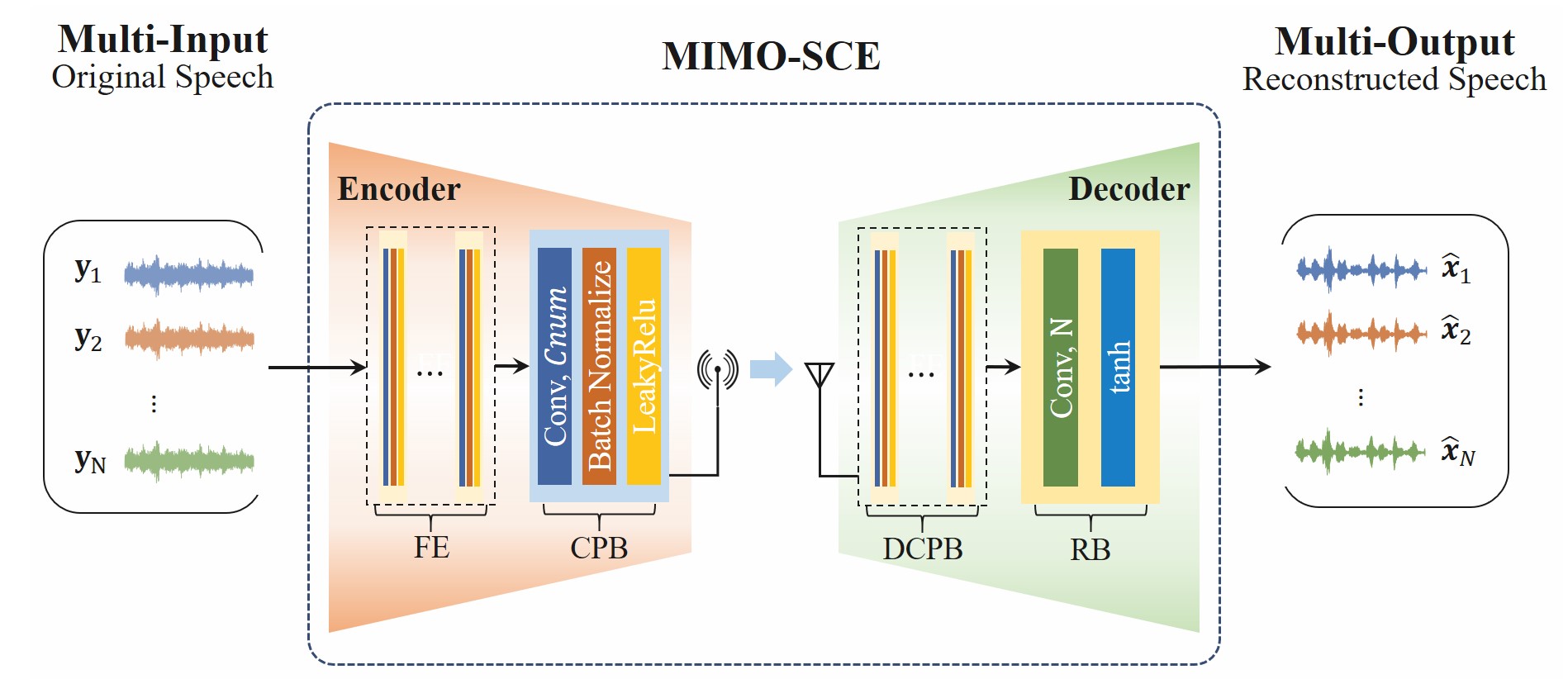}}
\caption{Architecture of MIMO-SCE\vspace{-0.1cm}}
\label{fig:MIMO-SCEN}
\end{figure}

\subsection{System architecture}
The proposed MIMO-SCE system is presented in Fig. \ref{fig:MIMO-SCEN}. The system comprises an encoder and decoder. During training, for the noisy multichannel input $\mathbf{Y}=[\mathbf{y}_{1},\mathbf{y}_{2},\cdots,\mathbf{y}_{N}]$, the MIMO system aims to generate enhanced speech signals $\hat{\mathbf{X}}$, where $\hat{\mathbf{X}}=f_{\theta}(\mathbf{Y}$). In addition, ${\theta}$ denotes the model parameters, which is estimated through Eq. \eqref{eq:training} by minimizing the difference between $\hat{\mathbf{X}}=[\hat{\mathbf{x}}_{1}, \hat{\mathbf{x}}_{2},\cdots,\hat{\mathbf{x}}_{N}]$ and the clean multichannel reference, $\mathbf{X}=[{\mathbf{x}}_{1}, {\mathbf{x}}_{2},\cdots,{\mathbf{x}}_{N}]$.
\begin{equation}
\begin{aligned}
\label{eq:training}
\hat{\theta} &= \mathop{\arg\min}_{\theta}D({f_{\theta}(\mathbf{Y}), \mathbf{X}}),
\end{aligned}    
\end{equation}
where $D(\cdot)$ denotes the cost function, which is defined in Eq. \eqref{eq:MSE}.
\begin{equation}\label{eq:MSE}
D(f_{\theta}(\mathbf{Y}), \mathbf{X}) =\mathop{\sum_{i=1}^{N}(\hat{\mathbf{x}}_{i}-\mathbf{x}_{i})^2}.
\end{equation}

After training, we place the encoder and decoder parts of the trained model at the edge and server sides, respectively. During the test, for the noisy multichannel input, the MIMO system first encodes the data into a latent representation with a reduced dimension. The encoded representation vectors are then transmitted to the server side and finally reconstructed to multichannel outputs based on the decoder. Because the latent representations (instead of original multichannel inputs) are transmitted, the data size is reduced; thus, online transmission bandwidth costs can be reduced.

\subsection{MIMO-SCE(F) and MIMO-SCE(S)}
The proposed MIMO-SCE(F) and MIMO-SCE(S) process speech signals in the time domain. The main advantage of time-domain speech signal processing is that the phase information can be more accurately preserved when compared with spectral-domain processing. 

For MIMO-SCE(F) and MIMO-SCE(S), we designed a bottleneck architecture, where the middle layer has few dimensions and is used to compress multichannel inputs. The middle layer was termed as the compression block (CPB). By assigning the filter number of the CPB to $C_{num}$, the compression rate is $R_{comp}=N/C_{num}$, which is derived from the channel number before and after the encoder. The inputs of MIMO-SCE(F) and MIMO-SCE(S) are the same as those used in the MISO systems, as shown in Figs. \ref{fig:MI-FCN} and \ref{fig:MI-SFCN}, respectively, and the outputs of the two systems are multichannel signals, as shown in Fig. \ref{fig:MIMO-SCEN}.

The MIMO-SCE(F) encoder consists of a feature inductor (FE) and CPB, where the FE is combined using four-layer CBs. All CBs have identical architectures, including Conv with filter number $fn=30$, filter length $fl=55$, batch normalization, and LeakyRelu. The CPB has a filter length of $fl=55$, filter number $C_{num}=1$, batch normalization, and LeakyRelu. The decoder consists of a decompression block (DCPB) and reconstruction block (RB). The DCPB also has four-layer CBs that decompress the transmission signal. The CB set is the same as the encoder. The RB has a Conv layer with a filter length $fl=55$, filter number $c=N$, and tanh activation function to rebuild the multichannel speech data, where $N=7$ in this study.

The encoder and decoder design of MIMO-SCE(S) is similar to that of MIMO-SCE(F). However, in MIMO-SCE(S), SincConv is added as the encoder’s first CB to extract additional speech features, as shown in Fig.\ref{fig:MI-SFCN}. The remainder of MIMO-SCE(S) is identical to MIMO-SCE(F).

For MIMO-SCE(F) and MIMO-SCE(S), we use the cost function in Eq. \eqref{eq:MSE} to estimate the model parameters. The compression rates of both systems were $7/1$.

\section{Experiments}
In this section, we introduce the experimental setup and results of the proposed MIMO-SCE framework. The compression ratio ($R_{comp}$) was maintained at $7$. The speech quality (measured via the perceptual evaluation of speech quality, PESQ \cite{rix2001perceptual}) and intelligibility (measured via short-time objective intelligibility, STOI \cite{taal2011algorithm}) of the enhanced multichannel outputs were measured and reported as the evaluation results. The PESQ score ranges from $0.5$ to $4.5$, and the STOI score typically ranges from 0 to 1. Higher PESQ and STOI scores indicate better speech quality and intelligibility, respectively.

\subsection{Experimental Setup}\label{sec:exp}
\begin{figure}[t]
\centering \centerline{
\includegraphics[width=0.4\textwidth,clip]{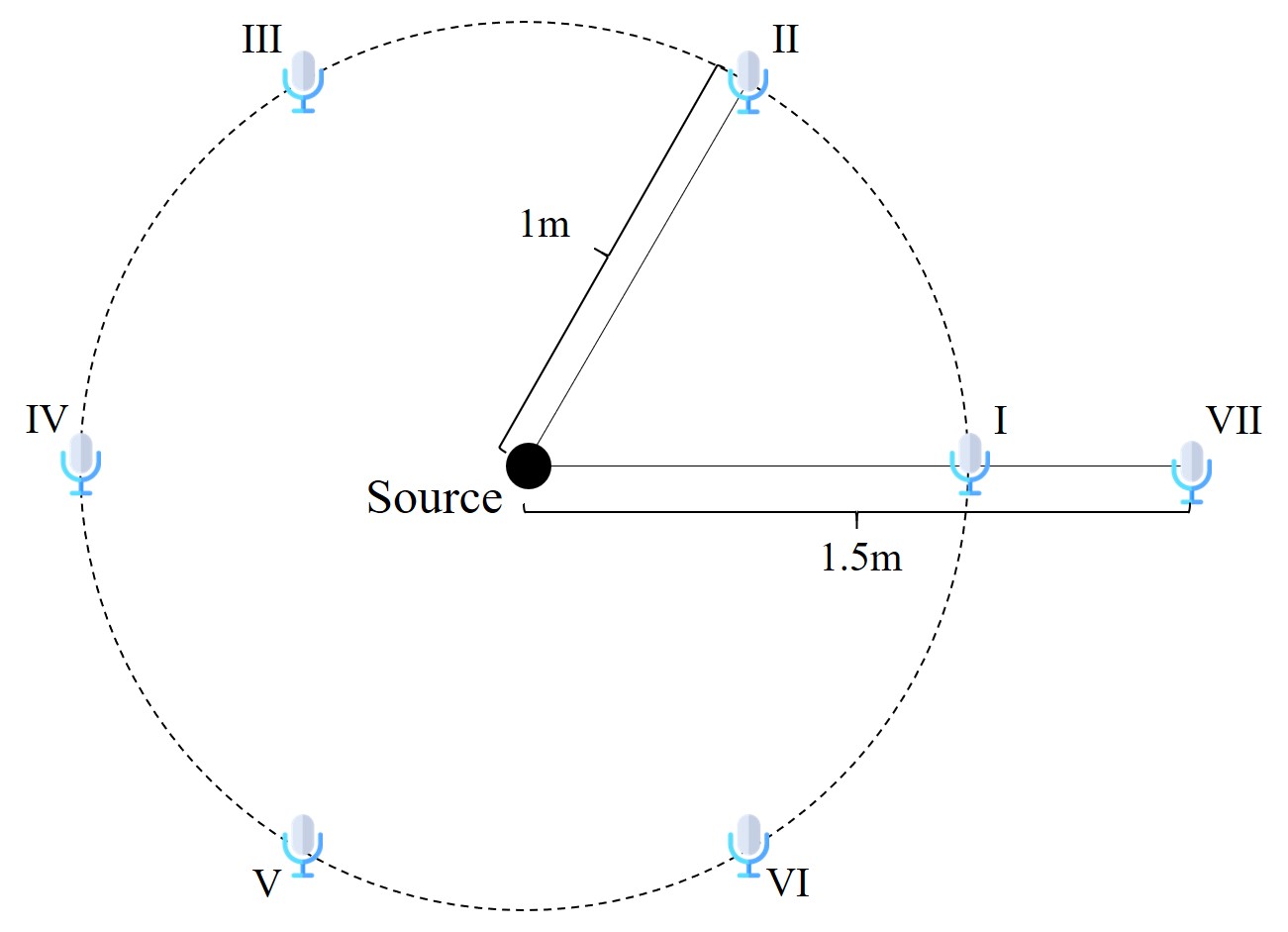}}
\caption{Recording settings for the experiments. The speaker is placed at the center (source) and surrounded by seven microphones (I to VII). Microphones I to VI are placed 1 m away from the source, whereas microphone VII is placed at 1.5 m and behind microphone I.}
\label{fig:mic}
\end{figure}

The speech data used in this study were recorded using the setup shown in Fig. \ref{fig:mic}. The speaker was placed at the center (Fig. \ref{fig:mic}) and surrounded by seven microphones. Six microphones—--I, II, III, IV, V, and VI—--were placed at a distance of 1 m from the source, whereas microphone VII was placed 1.5 m away from the source. All seven microphones were of the same model (Sanlux HMT-11). The transcript material is the Taiwan Mandarin Hearing in Noise Test dataset (TMHINT) \cite{wong2007development}, which is a phonetically balanced corpus consisting of 320 sentences and ten Chinese characters in each sentence. All utterances were pronounced by a native Mandarin male speaker and recorded at a sampling rate of 16kHz with seven microphones in a clean environment. We further split the 320 utterances into two parts: 250 for training and 70 for testing. The training utterances from the seven microphones were contaminated with eight noise types: pink, fan, babble, gun, alarm bell, cough, buccaneer, and engine, at signal-to-noise ratios (SNRs) of $-10$, $-5$, $0$, $5$ and $10$ dBs. Therefore, there are $35,000=250\times 7 \times 5\times 4$ noisy-clean utterance pairs in the training set. The testing utterances from the seven microphones were contaminated with another four noise types, namely sound of a water cooler, street noise, car noise, and the bell of a fire truck, at SNRs of $-10$, $-5$, $0$, $5$, and $10$ dB, consequently providing $9,800=70\times 7 \times 5\times 4$ noisy testing samples.

\subsection{Experimental results}
The qualitative and quantitative results of the proposed MIMO-SCE(F) and MIMO-SCE(S) are presented in this section. Those results of the testing noisy that denoted as ``Noisy'' in the following sections are also listed as the baseline.

\begin{figure}[t]
\begin{tabularx}{\columnwidth}{ m{0.5cm} m{3.2cm}m{4cm}} 
\toprule
 &  \:\:\:\:\:\:\:\:\:\:\:\:\:Waveform 
 &  \:\:\:\:\:\:\:\:\:Spectrogram \\ 
\midrule
(a) & \includegraphics[width=0.43\columnwidth]{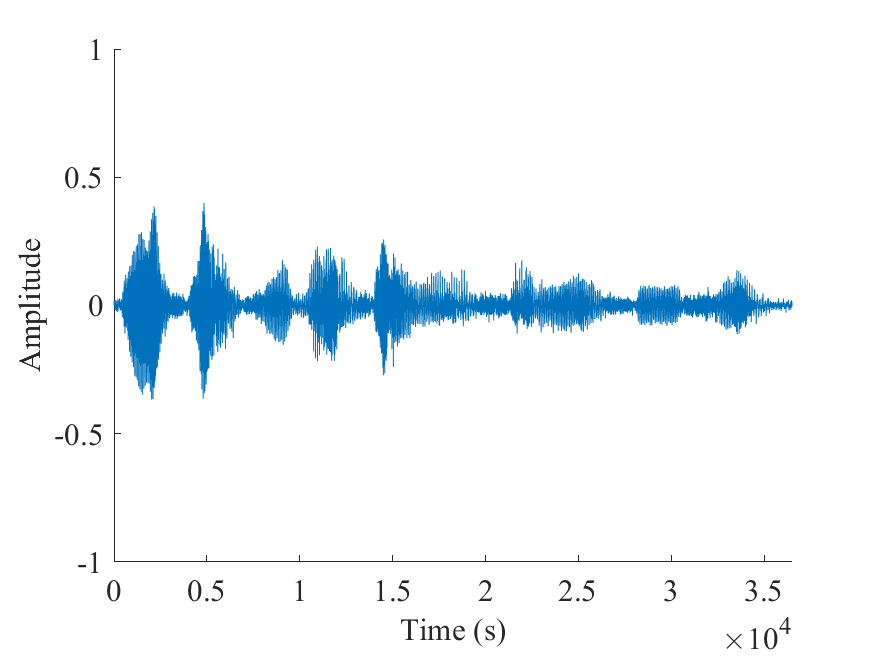} & \includegraphics[width=0.43\columnwidth]{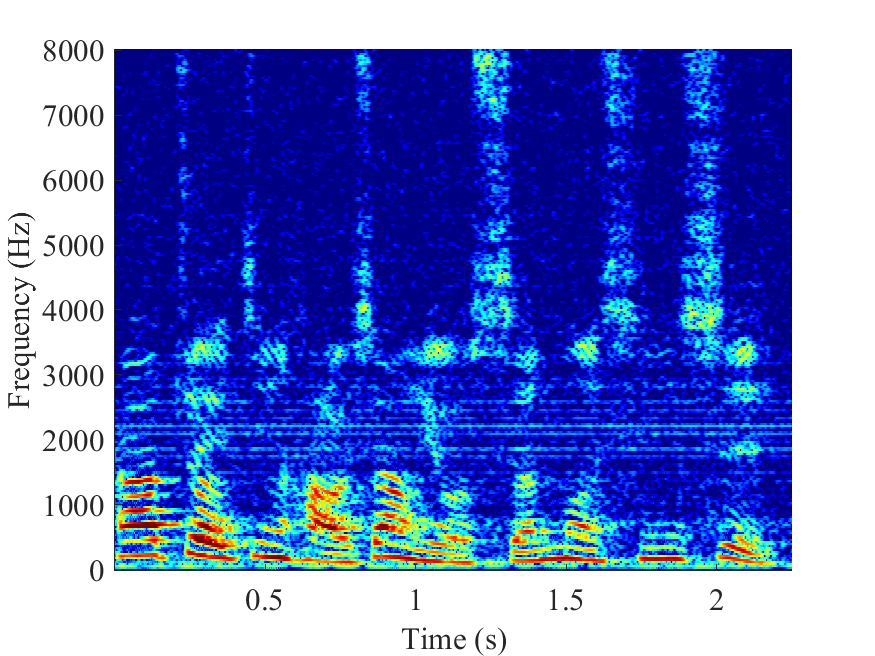}  \\ 
(b) & \includegraphics[width=0.43\columnwidth]{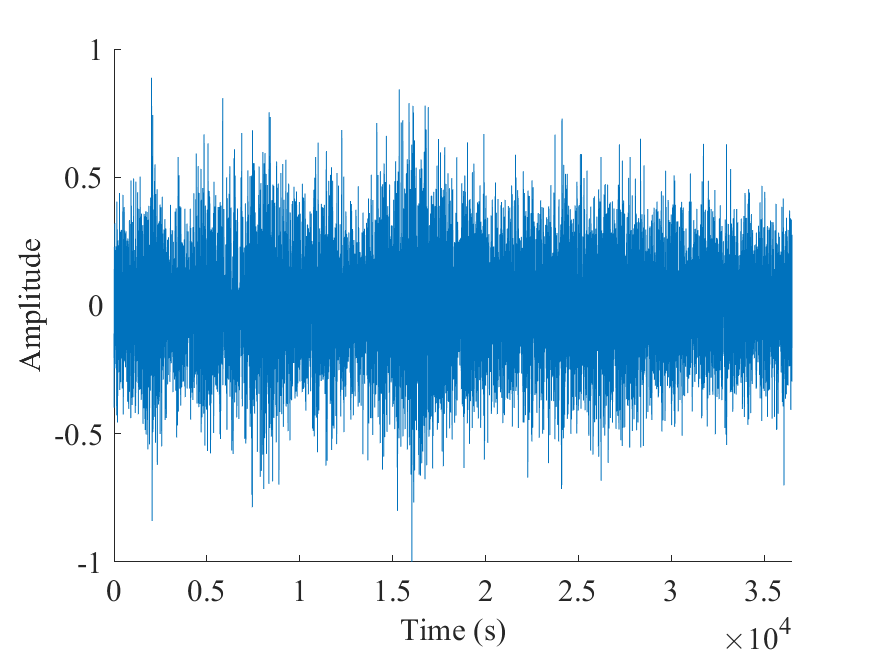} & \includegraphics[width=0.43\columnwidth]{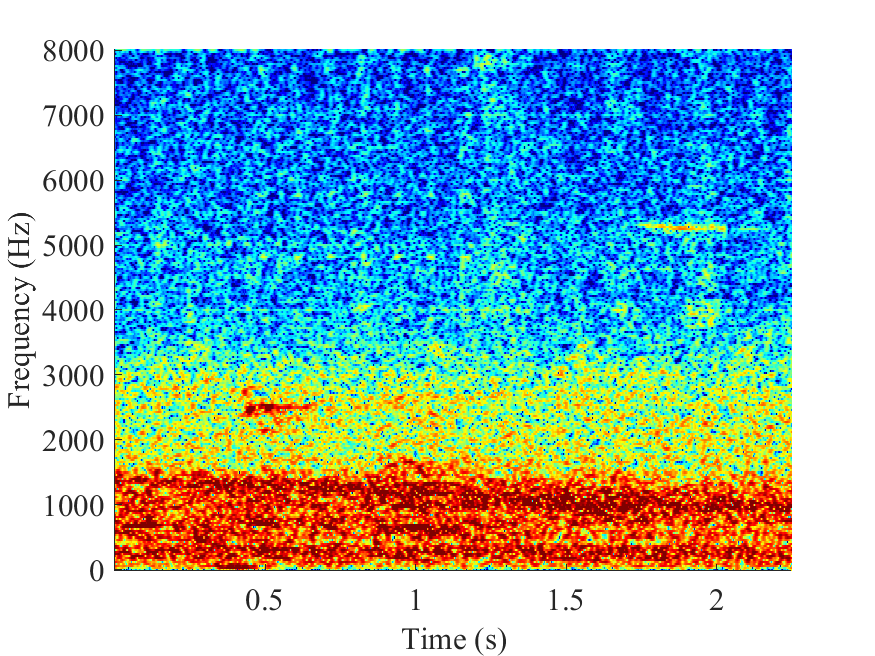}  \\ 
(c) & \includegraphics[width=0.43\columnwidth]{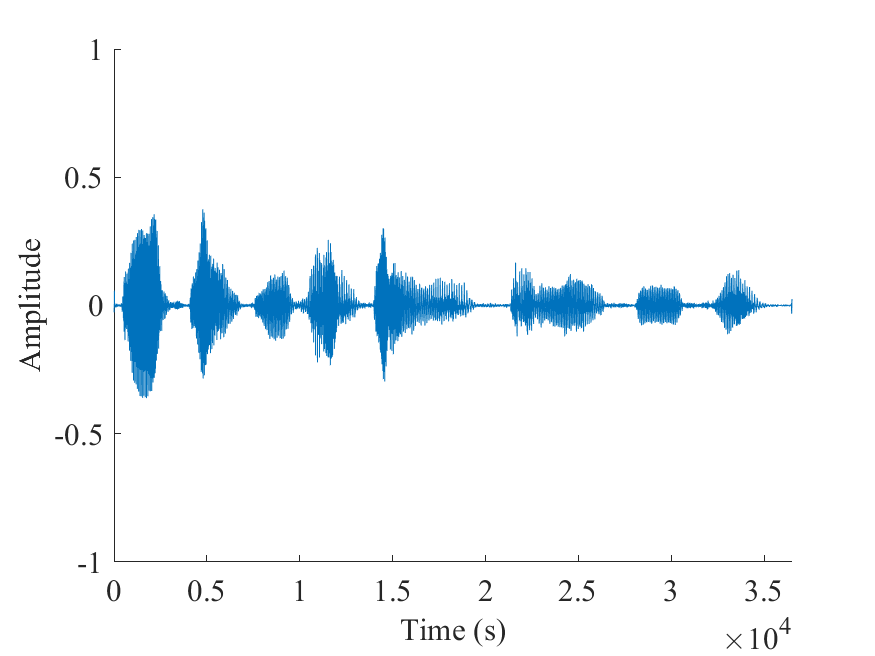} & \includegraphics[width=0.43\columnwidth]{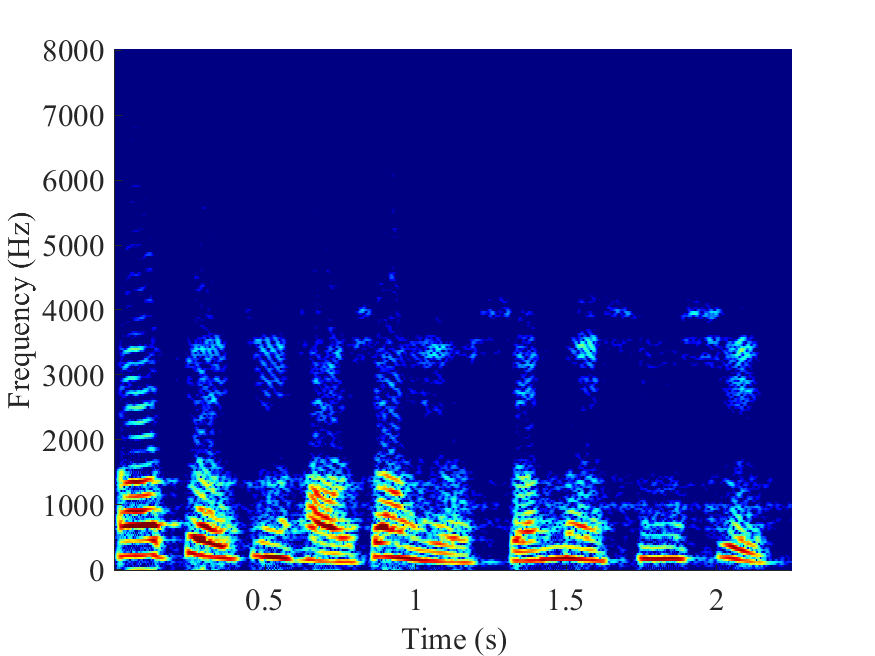}  \\ 
(d) & \includegraphics[width=0.43\columnwidth]{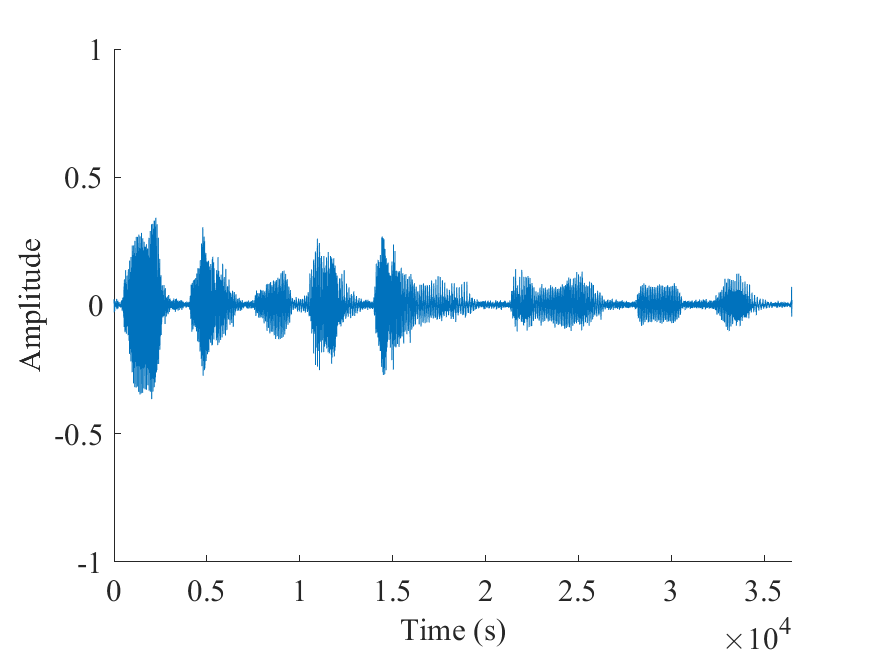} & \includegraphics[width=0.43\columnwidth]{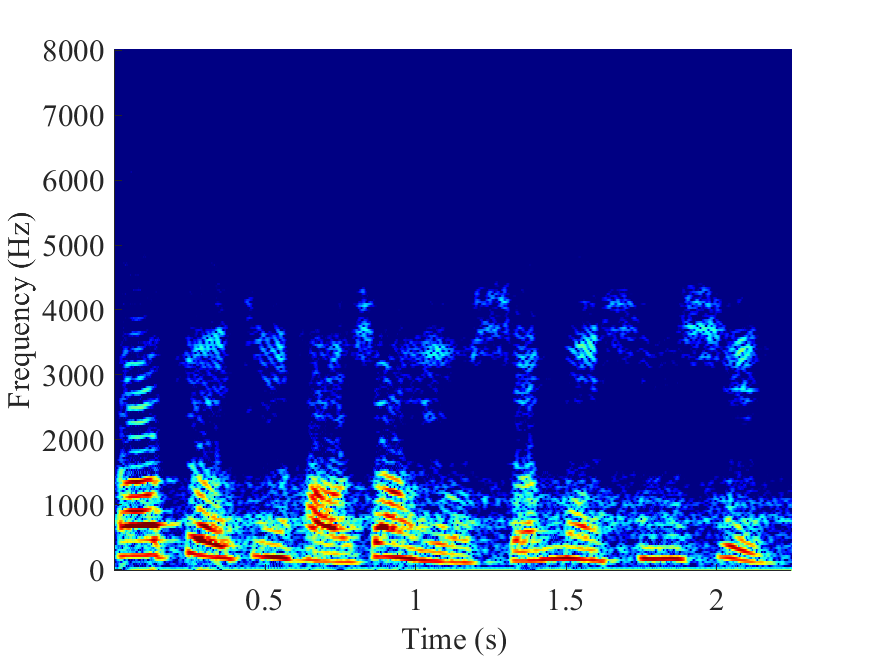}  \\ 
\bottomrule 
\end{tabularx}
\caption{Spectrogram and waveforms of (a) clean, (b) noisy (under street noise at 10 dB SNR), and (c) MIMO-SCE(S) with a compression ratio of 1, and (d) MIMO-SCE(S) with a compression ratio of 7. All utterances in the figure were selected from microphone III.}
\label{fig:wavsp}
\end{figure} 

\subsubsection{Qualitative spectrogram comparison}
We first demonstrate the effects of the compression ratio ($R_{comp}$) on the proposed MIMO-SCE(S), as depicted in Fig. \ref{fig:wavsp}, in terms of the spectrum plots and the associated waveforms of a sample utterance recorded from microphone III. Fig. \ref{fig:wavsp} (a) and (b) present the clean and noisy utterances, respectively, whereas (c) and (d) depict the utterances derived from MIMO-SCE(S) with compression ratios of 1 and 7, respectively. On comparing Figs. \ref{fig:wavsp} (c) and (d) with (b), it is evident that the noise components in the noisy spectrum and waveform were effectively suppressed. Furthermore, the harmonic structures of the spectrum and the envelope of the waveforms in Figs. \ref{fig:wavsp} (c) and (d) are preserved by MIMO-SCE(S), compared with those in Fig. \ref{fig:wavsp} (a). These results indicate the effectiveness of the proposed model in enhancing speech subjected to noise environments and a high compression ratio. Therefore, the MIMO-SCE models with a compression ratio of 7 were used and evaluated, as described in the following section. On the other hand, we also noted that the quality of the high-frequency components of the spectrum in Figs. \ref{fig:wavsp} (c) and (d) is degraded when compared with those of the clean spectra in Fig. (a). One possible inference is that handling highly distorted frequency-band signals remains a difficult task for the FCN model. 

\begin{table}[b]
\vspace{-0.2cm}\caption{Average PESQ and STOI scores of Noisy, MIMO-SCE (F), and MIMO-SCE(S)}
\label{tab:all_result}
\centering
\begin{tabularx}{\columnwidth}{c>{\centering}m{0.5cm}>{\centering}m{2cm}>{\centering}m{2cm}>{\centering\arraybackslash}m{2cm}}
\toprule
& & {\textbf{Noisy}} & {\textbf{MIMO-SCE(S)}} & {\textbf{MIMO-SCE(F)}}\\
\midrule
& \textbf{PESQ} & $1.825$ & $2.890$ & $2.927$ \\
& \textbf{STOI} & $0.678$ & $0.750$ & $0.801$ \\
\bottomrule
\end{tabularx}
\end{table}

\subsubsection{Quantitative objective evaluation results}
Table \ref{tab:all_result} lists the average PESQ and STOI results over the seven output channels and noise conditions for each of Noisy, MIMO-SCE(S) and MIMO-SCE(F). In addition, for each output channel, the associated clean signal in that channel was applied as a reference for performing PESQ and STOI metrics. From the table, we noted that MIMO-SCE(F) and MIMO-SCE(S) outperform Noisy in terms of the PESQ and STOI scores. MIMO-SCE(F) yields higher PESQ and STOI scores than MIMO-SCE(S), confirming the advantages of incorporating the SincConv layer in the enhancement system.

\begin{figure}[t]
\centering \centerline{
\includegraphics[width=1\columnwidth]{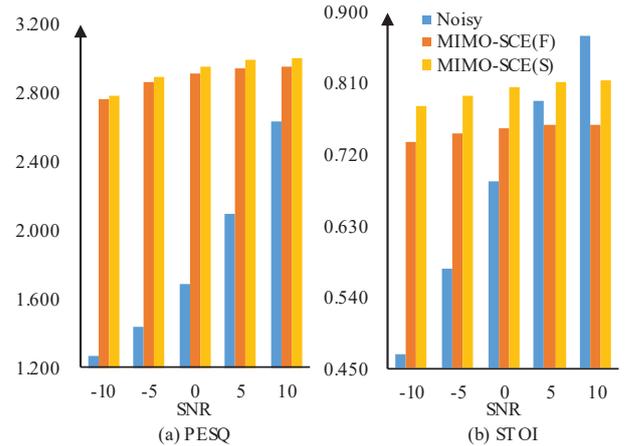}}
\caption{Averaged (a) PESQ and (b) STOI scores of Noisy, MIMO-SCE(F), and MIMO-SCE(S) in $-10$, $-5$, $0$, $5$ and $10$ SNRs.}
\label{fig:PESQ}
\end{figure}

To further analyze the results listed in Table \ref{tab:all_result}, we present the detailed PESQ and STOI scores of Noisy, MIMO-SCE(F), and MIMO-SCE(S) at specific SNRs ($-10$, $-5$, $0$, $5$ and $10$ dB) in Fig. \ref{fig:PESQ}. First, from Fig. \ref{fig:PESQ} (a), we note that both MIMO-SCE(F) and MIMO-SCE(S) improve PESQ scores over Noisy, and more significant improvements were observed at lower SNR levels. Meanwhile, MIMO-SCE(F) marginally outperforms MIMO-SCE(S) consistently over different SNR levels. From Fig. \ref{fig:PESQ} (b), we note that MIMO-SCE(F) and MIMO-SCE(S) improve the STOI scores over Noisy at low SNR conditions ($-5$ to $0$ dB); however, neither MIMO-SCE(F) nor MIMO-SCE(S) provide further enhancements over Noisy under cleaner conditions (at 5 and 10 dB SNRs). A possible cause of the reduced STOI scores is the distorted speech resulting from the data compression function of the proposed models.

\vspace{-0.1cm}\section{Conclusion}\label{sec:conc}
In this paper, we propose a novel MIMO-SCE system to perform data compression for the simultaneous transmission and enhancement of speech signals. We investigated two CDAE models, FCN and SFCN, as core models in the proposed system, with a short-hand notation MIMO-SCE(F) and MIMO-SCE(S), respectively. The experimental results show that, under a high compression ratio of 7, the proposed MIMO-SCE(F) and MIMO-SCE(S) models improve speech quality and reproducibility under various SNR conditions. To the best of our knowledge, this is the first study to simultaneously perform data compression and SE based on DL-based CDAE models in an MIMO scenario. In the future, we plan to explore the use of MIMO systems for the integration of other heterogeneous data, such as visual and textual data, to further improve data compression and SE efficacy. 
\bibliographystyle{IEEEtran}
\bibliography{mybib}

\end{document}